\documentclass[11pt,a4paper]{article}

\usepackage[T1]{fontenc}
\usepackage[utf8]{inputenc}
\usepackage{lmodern}
\usepackage[a4paper,margin=25mm]{geometry}
\usepackage{microtype}

\usepackage{amsmath,amssymb,mathtools}
\usepackage{booktabs,longtable,tabularx,array}
\usepackage{graphicx}
\usepackage{svg}
\usepackage{caption}
\usepackage{enumitem}
\usepackage{titlesec}
\usepackage{authblk}
\usepackage{xurl}
\usepackage{hyperref}

\hypersetup{
    colorlinks=true,
    linkcolor=blue,
    citecolor=blue,
    urlcolor=blue
}
\titleformat{\paragraph}[runin]
  {\normalfont\normalsize\bfseries}{}{0pt}{}
\titlespacing*{\paragraph}{0pt}{0pt}{0.6em}

\newcommand{\tableformat}{%
    \footnotesize
    \renewcommand{\arraystretch}{1.25}%
    \setlength{\extrarowheight}{2pt}%
}

\title{\textbf{Identifiability of degradation severities from condition-monitoring records: exact residual ambiguity under partial intervention labels}}

\author[1]{Ahmad BahooToroody}
\author[1]{Mohammad Mahdi Abaei}

\affil[1]{Department of Energy and Mechanical Engineering,
School of Engineering, Aalto University, Espoo, Finland}

\affil[]{\texttt{ahmad.bahoo@aalto.fi}}

\date{}

\begin{document}

\maketitle

\begin{abstract}
In many engineering systems, several degradation mechanisms can be active at the same time. These mechanisms may change the monitored signals in similar ways. The same measurements can then be explained by more than one combination of mechanisms. The required maintenance action, however, depends on which mechanism is present. This paper determines which mechanisms can be distinguished from the monitoring and intervention record when controlled interventions (seeded fault tests) label only some of them. Six acquisition conditions are assumed: (i) a healthy reference segment in every session; (ii) healthy data that separate operating-condition, session and noise effects from degradation; (iii) linearly independent mechanism signatures that remain constant with operating condition; (iv) verified health of every nontarget mechanism during each intervention; (v) at least one positive measured severity for each labelled mechanism; and (vi) sufficiently independent variation during passive operation. Under these conditions, Theorem 1 gives the complete set of mechanism decompositions that the record cannot distinguish. For three mechanisms, successive labels reduce the ambiguity dimension from nine to seven, five and three. When unlabelled mechanisms occur together, separate pairwise intensity bounds can be too restrictive or too permissive. In a two-mechanism example, cooler fouling is labelled and filter clogging remains unlabelled. The clogging trend is recoverable. The cooler inspection decision stays ambiguous until independent sensor calibration is added. A decision about a specific mechanism is justified when every explanation consistent with the record leads to the same outcome. An intensity trend for an unlabelled mechanism is guaranteed only when no other mechanism is unlabelled.
\end{abstract}


\section*{Nomenclature}

{\tableformat
\setlength{\tabcolsep}{3pt}

\begin{longtable}{@{}p{0.25\textwidth}p{0.71\textwidth}@{}}
\toprule
\textbf{Symbol, abbreviation, or term} &
\textbf{Engineering meaning} \\
\midrule
\endfirsthead

\toprule
\textbf{Symbol, abbreviation, or term} &
\textbf{Engineering meaning} \\
\midrule
\endhead

\bottomrule
\endfoot

PHM, FDI &
Prognostics and health management, and fault detection and isolation \\

MCR, NMF, semi-NMF &
Multivariate curve resolution, nonnegative matrix factorization and semi-NMF with one nonnegative factor \\

LL0--LL2 &
Label-ladder hypotheses: anchored deconvolution, intervention family and spanning \\

\(x,p,w,u\) &
Sensor vector, number of sensors, measured operating condition and acquisition session \\

\(K,L,U,q\) &
Number of mechanisms, labelled and unlabelled sets, and number \(q=|U|\) of unlabelled mechanisms \\

cell &
Observations sharing session, operating condition, intervention status and measured severity \\

healthy anchor &
Cell with all mechanism intensities zero \\

verified absence &
Independent confirmation that a nontarget mechanism is healthy \\

\(a_k,A\) &
Mechanism sensor signature and signature matrix \\

\(z_k,z\) &
Nonnegative mechanism intensity and intensity vector \\

\(l,h_k(l)\) &
Measured physical severity and corresponding intensity response \\

\(g(w),d,B\) &
Operating condition basis, dimension and sensor coefficients \\

\(c_u,\varepsilon\) &
Session offset and common measurement noise \\

\(S\) &
Passive intensity support in reference coordinates \\

\(\mathcal{M}\) &
Declared mechanism model class satisfying the acquisition structure \\

\(T,A',z'\) &
Invertible intensity coordinate change and alternative signatures and intensities \\

\(=_{d}\) &
Equality in distribution \\

\(R_L^S,\mathcal{T}_L\) &
Ambiguity set of transformations consistent with the record, and its superset preserving labelled intervention directions up to positive scale \\

\(\Gamma_S\) &
Coefficient vectors keeping transformed intensity nonnegative on passive support \\

\(D,H,N\) &
Labelled scaling, absorption into labelled intensities and unlabelled mixing blocks \\

\(\alpha_k,\gamma\) &
Positive intensity scale factors \\

\(\Phi_k(\alpha),b,\kappa_{kj}\) &
Absorption coefficients, vector and reference relative lower intensity ratio \\

\(GL_r,D_+^r\) &
Invertible real and positive diagonal matrices of order \(r\) \\

\(e_k,I,A^+\) &
Coordinate unit vector, identity matrix and left pseudoinverse \\

\(\mathrm{col},\mathrm{row},\mathrm{span},\mathrm{proj}_U\) &
Column space, row space, linear span and unlabelled coordinate restriction \\

\(Y,Z,Q_L,C,V\) &
Noiseless signal and intensity matrices, projection, projected signatures and coefficients \\

\(c,c_0,\rho_c\) &
Sensor index, unaffected sensor and independently calibrated signature bound \\

\(t,F,\overline{t},\Delta\) &
Shear coefficient, feasible set, calibration centre and calibration half width \\

\end{longtable}
}

\setcounter{table}{0}

\section{Introduction}

Condition monitoring systems across engineering sectors share a basic problem. Different degradation mechanisms may produce overlapping changes in measured variables, while the required maintenance action depends on which mechanism is present. In aircraft gas turbines, monitored engine variables are used to distinguish faults in different gas path components \cite{ref1}. In nuclear power plants, condition monitoring similarly uses several process measurements to distinguish faults in individual components \cite{ref2}. Maritime machinery provides another example. Cooler fouling and filter clogging can both alter monitored temperatures and pressures, although one requires cooler cleaning and the other filter replacement. The reliability question is therefore whether the available record can identify the degradation mechanism that determines the required inspection or repair. Coraddu et al.~\cite{ref3} investigated machine learning approaches to condition-based maintenance of naval propulsion plants. When alternative explanations cannot be separated, the record cannot justify repair directed at a specific mechanism, so maintenance may have to address the wider unit. PHM uses monitoring information to assess health and support maintenance decisions \cite{ref4}. For decisions concerning a specific mechanism, the key question is whether the available monitoring and intervention record can distinguish that mechanism before a fitted model is used to justify targeted maintenance.

Relevant studies can be grouped into three strands. Multisensor prognostics address how degradation is represented for prognosis. Song and Liu \cite{ref5} estimated a composite health index. Fang et al.~\cite{ref6} combined functional scores with regression for incomplete multisensor records. Coraddu et al.~\cite{ref3} used a naval propulsion record and its released benchmark \cite{ref7}, while Chehade et al.~\cite{ref8} estimated failure thresholds from sensor data. Compound fault separation addresses overlapping fault signatures. Wang et al.~\cite{ref9} combined empirical mode decomposition with separation based on statistical independence. Wang et al.~\cite{ref10} constrained NMF for compound bearing faults. Hao et al.~\cite{ref11} separated and prognosed degradation caused by multiple mechanisms under an assumption of identical mechanisms. Structural fault isolability asks which faults a model and sensor set can distinguish. Murata et al.~\cite{ref12} supplied structural residual generators for FDI, and Düştegör et al.~\cite{ref13} analysed structural fault isolability on a benchmark process. Fu et al.~\cite{ref14} considered unknown and unlabelled failure modes under one dominant mode rather than several mechanisms acting at the same time.

These studies typically produce a prognostic estimate, separated fault signals, or a fault isolation design based on a model \cite{ref5,ref6,ref9,ref10,ref12}. Song and Liu \cite{ref5} also addressed scale ambiguity in a health index model, showing that estimating an index and assigning its physical interpretation are distinct tasks. The question here is different. For a partially labelled intervention record, what is the complete family of signature and intensity assignments that could reproduce the same observations? The studies above select one member of this family through their estimation choices \cite{ref5,ref6,ref9,ref10,ref12,ref14}. For maintenance directed at a specific mechanism, however, the engineer must know whether all other explanations consistent with the record lead to the same mechanism assignment.

Reliability theory already distinguishes an observable failure record from the hidden quantities it can determine. Tsiatis \cite{ref15} demonstrated nonidentifiability in competing risks. Bedford and Lindqvist \cite{ref16} treated identifiability for repairable systems under competing risks. Massonis and Villaverde \cite{ref17} and Massonis et al.~\cite{ref18} used symmetry and reparameterisation to determine which quantities remain unchanged in dynamic models. Tsiatis \cite{ref15}, for example, showed that the same observed information on competing risks can be associated with different joint distributions of potential failure times. Condition monitoring poses the same type of identifiability question with different observations. Here the data are sensor signals and intervention records. The hidden quantities are the sensor signatures assigned to degradation mechanisms and their nonnegative intensities. A mechanism quantity is recoverable only when every explanation consistent with the record assigns it the same value.

Lawton and Sylvestre \cite{ref19}, Tauler et al.~\cite{ref20}, and Borgen and Kowalski \cite{ref21} established rotational ambiguity, selective regions and feasible region construction in MCR. Here, rotational ambiguity means that different combinations of hidden mechanism signatures and their contributions can reproduce the same measured signal. Rajkó \cite{ref22} described the nonnegativity of one factor using polar or dual cone geometry. Golshan et al.~\cite{ref23,ref24} computed and reviewed feasible sets for three profiles. Beyramysoltan et al.~\cite{ref25} showed how equality constraints reduce the feasible set. Manne \cite{ref26} showed that selective samples fix signature directions, while Sawall et al.~\cite{ref27} and Neymeyr and Sawall \cite{ref28} derived results for complementary spaces and coupling. Feasible factor sets and dual Borgen plots \cite{ref29,ref30,ref31}, feasible solutions \cite{ref32}, and partial column uniqueness \cite{ref33} further restrict the possible rotational families. Donoho and Stodden \cite{ref34}, Huang et al.~\cite{ref35}, and Ding et al.~\cite{ref36} study separable, scattered or semi-NMF settings. In the engineering acquisition considered here, controlled interventions provide the selective observations, mechanism intensities remain nonnegative, and mechanism signatures may contain both positive and negative sensor responses.

Therefore, this paper aims to determine which mechanism explanations an engineering condition monitoring record leaves indistinguishable when intervention sessions identify only a prescribed subset of degradation mechanisms. The formulation is not tied to a particular industrial sector. Maritime machinery is used as the running example. The analysis combines signature directions fixed by interventions with nonnegative intensities over passive occurrence patterns. This use of intervention design to determine which model features can be identified is related to qualitative experiment design as described by Walter and Pronzato \cite{ref37}. Sawall et al.~\cite{ref38} apply the idea to ambiguity in curve resolution, while Escobar and Meeker \cite{ref39} and Meeker and Escobar \cite{ref40} treat multiple stress levels and separated degradation factors as requirements in accelerated testing.

The acquisition hypotheses LL0--LL2 are called the label-ladder hypotheses. The name refers to the effect of the intervention labels: each additional label is one rung of a ladder, and each rung removes exactly \(K-1\) ambiguity dimensions (Corollary 2, Table~\ref{tab:labelsets}). The ladder has three levels. LL0 (anchored deconvolution) separates degradation from operating-condition, session and noise effects. LL1 (intervention family) states what each controlled intervention must supply. LL2 (spanning) requires sufficiently independent variation during passive operation. Under LL0, LL1 and LL2, Theorem 1 describes all indistinguishable decompositions using \(R_L^S\). Only quantities that remain constant across this set are recoverable, regardless of the estimator. Buchholz et al.~\cite{ref41}, Squires et al.~\cite{ref42} and Ahuja et al.~\cite{ref43} use interventions for causal representations. In this paper, the intervention sessions instead identify named mechanism directions and verify that nontarget mechanisms are absent.

The example with three mechanisms shows how intervention coverage changes the number of free parameters and the quantities that remain recoverable. Figure~\ref{fig:acquisition} connects the acquisition requirements to the allowed explanations, and Table~\ref{tab:labelsets} compares different label sets. Proposition 1 and Figure~\ref{fig:pairwise} show why joint occurrence matters. Separate pairwise bounds can reject an allowed assignment or retain one that is not allowed. The example with two mechanisms then considers independent sensor calibration using the idea of net analyte signal calibration \cite{ref44}. Figure~\ref{fig:calibration} shows the two roles of calibration by illustrating how sensor calibration restricts mechanism mixing and how intervention sessions calibrate the severity response. These results support acquisition design by showing when another labelled intervention, a wider range of passive operating states, or independent calibration is needed for a decision concerning a specific mechanism. Table~\ref{tab:prior} relates each result of this paper to the closest prior result and states what is new in each case.

{\footnotesize
\renewcommand{\arraystretch}{1.08}
\setlength{\extrarowheight}{0.5pt}
\setlength{\LTpre}{0.45em}
\setlength{\LTpost}{0.45em}

\begin{longtable}{@{}p{0.21\textwidth}p{0.23\textwidth}p{0.50\textwidth}@{}}
\caption{Relation of the present results to prior ambiguity and identifiability results.}
\label{tab:prior}\\
\toprule
\textbf{Result in this paper} &
\textbf{Closest prior result} &
\textbf{New element here} \\
\midrule
\endfirsthead

\toprule
\textbf{Result in this paper} &
\textbf{Closest prior result} &
\textbf{New element here} \\
\midrule
\endhead

\bottomrule
\endfoot

Theorem 1: exact ambiguity set under partial labels &
Selectivity and resolution \cite{ref20,ref26} &
Verified absences, session and nuisance model, sign-unrestricted signatures, and an exact characterization that also covers the severity responses \\[1.0em]

Corollary 1: invariant spans and block form &
Complementarity and coupling \cite{ref27,ref28} &
Block transformation under label constraints; no sign restriction on the signatures (Appendix A) \\[1.0em]

Corollary 2: dimension \(K^2-|L|(K-1)\) &
Equality constraints \cite{ref25}; known zeros \cite{ref45} &
Explicit count under direction-plus-absence labels; passive nonnegativity does not reduce the dimension \\[1.0em]

Corollary 3: severity-response ratios &
Accelerated-test planning \cite{ref39,ref40} &
Identifiability limited to recorded levels under an unknown monotone response \\[1.0em]

Corollary 4: single unlabelled scale &
Partial identifiability of NMF \cite{ref33} &
Semi-NMF with intervention labels; preservation of the intensity preorder \\[1.0em]

Proposition 1: pairwise bounds fail &
Feasible regions \cite{ref23,ref24} &
Explicit counterexamples to necessity and to sufficiency under joint occurrence \\[1.0em]

Decision sufficiency (Definition 1) &
Set identification in econometrics \cite{ref46,ref47} &
Formulation for maintenance rules and a linear-programming check \\[0.2em]

\end{longtable}
}
\section{Condition monitoring record and hypotheses}

Consider an engineering system monitored during controlled intervention sessions and passive operation. In an intervention session, one degradation mechanism is deliberately introduced or varied. During passive operation, mechanisms are observed without imposed labels and may occur separately or together. Maritime machinery provides the running example, with cooler fouling and filter clogging. The analysis concerns the population record: the record that would be observed, under the stated acquisition, if unlimited data were available. It refers to one monitored system, not to a fleet of machines.

The measured sensor vector \(x\in\mathbb{R}^{p}\) is represented as the sum of degradation, measured operating condition effects, an offset specific to each session and measurement noise. This representation separates degradation from changes caused by load, speed, ambient condition or acquisition session. With measured condition \(w\in\mathbb{R}\) and session index \(u\),

\begin{equation}
x=Az+B\,g(w)+c_u+\varepsilon,
\qquad
A=
\begin{bmatrix}
a_1 & \cdots & a_K
\end{bmatrix}.
\label{eq:model}
\end{equation}

Here, \(a_k\) is the mechanism signature, which is the fixed sensor response pattern produced by mechanism k. Its intensity \(z_k\geq0\) is nonnegative and increases strictly with physical severity. The declared condition basis \(g:\mathbb{R}\rightarrow\mathbb{R}^{d}\) and its coefficients \(B\in\mathbb{R}^{p\times d}\) represent the measured operating condition effect. The term \(c_u\in\mathbb{R}^{p}\) represents an offset specific to each acquisition run. The common noise \(\varepsilon\) is independent of mechanism intensity and operating condition. The signature matrix \(A\in\mathbb{R}^{p\times K}\) may contain both positive and negative entries.

A session is one acquisition run. A cell groups observations that share the same session, operating condition, intervention or healthy status, and measured severity. For each observation, the record also contains the measured sensor vector \(x\). A healthy anchor is a cell in which all mechanism intensities are zero. An intervention label identifies the mechanism that was deliberately introduced or varied. Measured achieved severity is the independently measured physical severity of that labelled mechanism. A verified absence means that a nontarget mechanism has been independently confirmed as healthy. Passive records have no mechanism labels. Their possible intensity combinations define the pooled passive support \(S\). Let \(\mathcal{K}=\{1,\ldots,K\}\) index all mechanisms, \(L\subseteq\mathcal{K}\) the labelled set, \(U=\mathcal{K}\backslash L\) the unlabelled set, and \(q=|U|\). The record preserves the identity of each labelled mechanism, so labelled mechanisms cannot be permuted within \(L\).

LL0 establishes whether healthy reference data can separate operating condition effects, session effects and measurement noise from the degradation signal.

\paragraph{LL0 (anchored deconvolution).}
(a) Every analysed session contains a healthy anchor, that is, a cell in which all mechanism intensities are zero. (b) The declared basis \(g\) has no intercept component. Some sessions contain two or more healthy anchors at different operating conditions. For such a session, form the differences \(g(w)-g(w')\) between its anchor conditions. These differences, collected over all sessions, span \(\mathbb{R}^{d}\). (c) The common noise \(\varepsilon\) has an everywhere-nonzero characteristic function, and its location is fixed by setting the lower median of each sensor coordinate to zero.

The lower median of a coordinate \(\varepsilon_c\) is the smallest value \(m\) with \(P(\varepsilon_c\leq m)\geq1/2\). It is used because it is uniquely defined even when the median is not unique. Any other fixed location measure, for example the mean when it exists, would serve equally. The everywhere-nonzero characteristic function allows deconvolution: division by the noise characteristic function in the Fourier domain is then well defined. Gaussian noise and most common noise models satisfy this condition.

\paragraph{Identification under LL0.}
Within one session the offset \(c_u\) is constant. The difference between two healthy-anchor medians therefore equals \(B(g(w)-g(w'))\). Because these differences span \(\mathbb{R}^{d}\) by condition (b), they determine \(B\). A healthy anchor then determines \(c_u\) in each session. After \(B\,g(w)+c_u\) is removed, the healthy cells give the noise distribution. Condition (c) then makes the deconvolution of every cell unique, which gives the distribution and support of \(Az\) in every cell. The pooled passive support \(S\) is expressed in the coordinates of a reference signature matrix \(A\), with \(S=A^{+}\,\mathrm{supp}(Az)\). The matrix \(A^{+}\) is its left pseudoinverse. All statements concern the population model in the declared model class \(\mathcal{M}\).

LL1 specifies the information required from controlled intervention sessions.

\paragraph{LL1 (intervention family).}
There are \(K\) mechanisms and one shared condition axis. The signature matrix \(A\) has full column rank, \(\mathrm{rank}\,A=K\), and remains constant with operating condition \(w\). For each labelled mechanism \(k\in L\), interventions involving only that mechanism occur at one or more positive measured severities \(l>0\). Each such intervention cell is point supported at \(h_k(l)e_k\), meaning that the cell has one fixed intensity vector. Two or more distinct positive severities are needed only for the normalized severity response in Corollary 3. The response map \(h_k\) is unknown but strictly increasing and satisfies \(h_k(0)=0\). Therefore, positive measured severity corresponds to positive mechanism intensity. During each intervention, every nontarget mechanism \(j\neq k\) is independently verified as healthy, giving \(K-1\) verified absences per intervention. Passive operation has pooled support \(S\subseteq\mathbb{R}_{\geq0}^{K}\).

\paragraph{LL2 (spanning).}
LL2 requires enough independent variation during passive operation to represent all remaining unlabelled mechanism directions. The reference satisfies

\begin{equation}
\mathrm{span}\!\left(\mathrm{proj}_{U}S\right)
=
\mathbb{R}^{q},
\qquad
\text{equivalently}
\qquad
\mathrm{span}\!\left(S\cup\{e_k:k\in L\}\right)
=
\mathbb{R}^{K}.
\label{eq:LL2}
\end{equation}

When the pooled passive second moment matrix \(\mathbb{E}[z_Uz_U^{\top}]\) exists and is finite, with positive weight on every passive cell, it must be nonsingular. In other words, the passive intensities must vary in sufficiently independent directions. The main statement does not require this moment matrix to be finite. When \(q=0\), LL2 places no additional requirement. When \(q=1\), LL2 is automatically satisfied once the unlabelled mechanism becomes active.

The declared model class \(\mathcal{M}\) contains parameterizations satisfying LL1 after LL0 has fixed the nuisance terms. Lemma A.1 shows that any two observationally equivalent models in \(\mathcal{M}\) share \(B\), the offsets \(c_u\) and the noise law. These nuisance quantities can therefore be fixed without loss of generality, and equivalence reduces to the pair \((A,z)\). The intensity distributions within individual cells are otherwise unrestricted. Two explanations are equivalent when they reproduce every sensor distribution and intervention record. A quantity is recoverable when it has the same value in every equivalent explanation. Alternative intensities must remain nonnegative for every passive state. The set of coefficient vectors that satisfy this requirement is called the support dual cone

\begin{equation}
\Gamma_S
=
\left\{
v\in\mathbb{R}^{K}:
v^{\top}z\geq0
\ \text{for every }z\in S
\right\}.
\label{eq:dualcone}
\end{equation}

A coefficient vector belongs to this cone only if it keeps the transformed intensity nonnegative for every passive state. Write \(GL_K\) for the invertible real \(K\times K\) matrices, \(D_+^r\) for the positive diagonal matrices of order \(r\), and
\[
\mathcal{T}_L
=
\left\{
T\in GL_K:
Te_k=\alpha_ke_k,\;
\alpha_k>0,\;
k\in L
\right\}.
\]
The dimension counts the number of continuously adjustable parameters that remain after the label constraints are imposed. Formally, it is the topological dimension within the constrained linear space.

\begin{table}[!htbp]
\centering
\caption{Acquisition requirements and possible engineering implementations. Maritime examples are illustrative. The requirements apply to any engineering system satisfying LL0, LL1 and LL2.}
\label{tab:acquisition}

{\footnotesize
\renewcommand{\arraystretch}{1.30}
\setlength{\extrarowheight}{2pt}

\begin{tabularx}{\textwidth}{@{}p{0.19\textwidth}p{0.37\textwidth}X@{}}
\toprule
\textbf{Requirement} &
\textbf{Possible engineering implementation} &
\textbf{Evidence boundary} \\
\midrule

LL0 anchor in every session &
Record known healthy operation within each analysed run, before seeding or after verified restoration. &
A separate healthy reference campaign cannot supply another session's anchor. \\[3.0em]

LL0 condition and noise calibration &
Vary the declared condition during healthy segments and record session identity and repeated measurements. &
The model still requires a basis without an intercept, anchor differences that span the condition space and common noise that can be deconvolved. \\[3.0em]

LL1 measured achieved severities &
For fouling, measure retained deposit mass independently. For clogging, measure blocked filter area at stabilized levels. &
Severity definitions require a justified monotone response. Actuator settings alone are insufficient. \\[3.0em]

LL1 nontarget absences &
Inspect or independently instrument each nontarget mechanism and record its healthy status in every intervention session. &
Scenario names and binary anomaly flags do not certify these absences. \\[3.0em]

LL1 signatures and LL2 passive variation &
Check signature constancy within the declared regime and establish spanning unlabelled variation during passive operation. &
More samples, load variation or sensor rank alone do not certify these assumptions at the mechanism level. \\[0.5em]

\bottomrule
\end{tabularx}
}
\end{table}

A derived measure such as temperature effectiveness or pressure drop can represent physical severity only when it has been calibrated independently for the relevant operating regime. The calibration must also separate the effects of nontarget mechanisms. Reusing a sensor signal that is already part of the mixed response being decomposed does not provide independent severity information.

\begin{figure}[!htbp]
\centering
\includegraphics[width=0.86\linewidth]{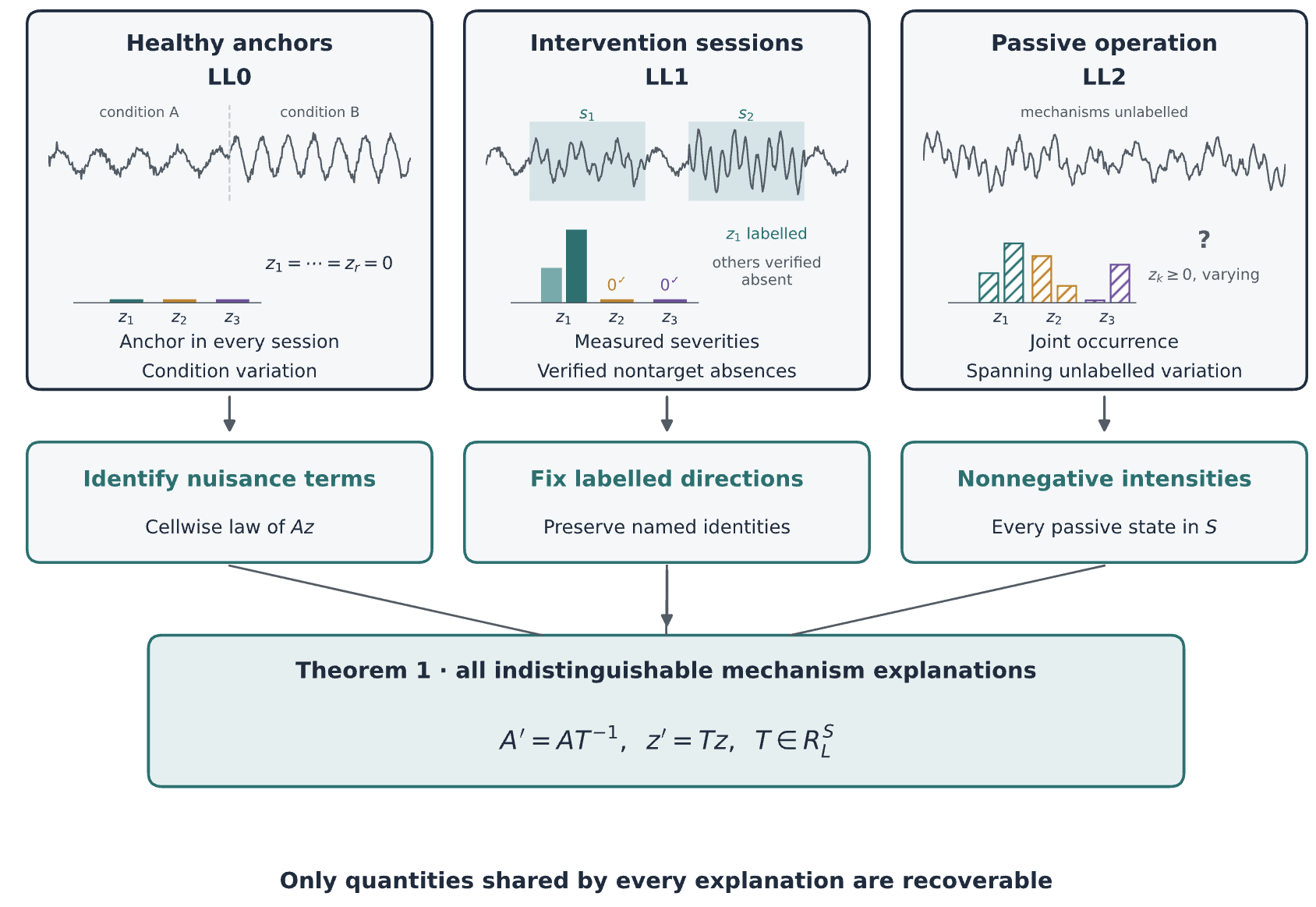}
\caption{Acquisition structure and analytical sequence for assessing mechanism recoverability at the population level. LL0, LL1 and LL2 provide different constraints whose intersection retains every explanation allowed by the model. Estimators and priors may select within this set through additional assumptions or preferences. Such selection alone does not extract further identifying information from the record.}
\label{fig:acquisition}
\end{figure}

Figure~\ref{fig:acquisition} summarizes the three roles. Healthy anchors separate degradation from nuisance effects. Intervention sessions fix the directions of named mechanisms. Passive operation restricts the remaining intensity assignments that can stay nonnegative. Their intersection defines the explanations considered by Theorem 1.

\section{Mechanism explanations consistent with the monitoring record}

This section describes the ambiguity set, which contains all alternative signatures and intensities that reproduce the same intervention and passive operation record.

\paragraph{Theorem 1 (exact ambiguity set).}
Under LL0, LL1 and LL2, the set of equivalent signature matrices is exactly \(\{AT^{-1}:T\in R_L^S\}\), where

\begin{equation}
R_L^S
=
\left\{
T\in\mathcal{T}_L:
\mathrm{row}_i(T)\in\Gamma_S,\;
i\in\mathcal{K}
\right\}.
\label{eq:ambiguity}
\end{equation}

For every equivalent explanation, the intensity distribution in each cell is transformed by \(z'=Tz\), with \(h'_k(l)=\alpha_kh_k(l)\) at recorded severities. Conversely, these transformations reproduce the record when the unobserved severity responses are extended monotonically as required by LL1. The set \(R_L^S\) is closed in \(GL_K\).

\paragraph{Proof.}
First, LL0 identifies the distribution of \(Az\) in every cell. LL1 supplies the direction of each labelled mechanism, while Eq.~\eqref{eq:LL2} supplies the remaining independent variation from passive operation. Together, the labelled intervention directions and the independent passive variation span \(\mathrm{col}\,A\). Any equivalent rank \(K\) signature matrix \(A'\) must therefore have the same column space. Hence \(T=A'^{+}A\) is invertible and satisfies \(A'T=A\). Both signature matrices have full column rank. Therefore, each degradation signal corresponds to only one intensity vector, and the alternative cellwise distribution must satisfy \(z'=_{d}Tz\).

Second, at each positive recorded severity, every equivalent explanation must retain the named mechanism and the recorded absences. Hence \(Te_k=\alpha_ke_k\), with \(\alpha_k>0\). This requirement fixes each labelled transformation direction up to a positive scale. During passive operation, every alternative intensity must remain nonnegative for every state in \(S\). This condition places each row of \(T\) in \(\Gamma_S\). Therefore, every equivalent explanation must satisfy Eq.~\eqref{eq:ambiguity}.

Third, consider any transformation satisfying Eq.~\eqref{eq:ambiguity}. Set \(A'=AT^{-1}\), transform each intensity distribution using \(T\), and prescribe \(h'_k(l)=\alpha_kh_k(l)\) at recorded severities. The fixed labelled directions preserve the intervention labels and recorded absences. The row constraints preserve nonnegative intensities during passive operation. LL1 allows monotone extensions at unobserved severities. Because \(A'z'=Az\), the complete observable record is reproduced. Each row constraint belongs to the closed cone in Eq.~\eqref{eq:dualcone}. Within \(GL_K\), a labelled scale cannot reach zero because this would make \(T\) singular. Therefore, the defining constraints are closed relative to the stated matrix space, and the ambiguity set is closed there.

The necessity and converse parts of the proof show that Eq.~\eqref{eq:ambiguity} contains exactly the explanations that reproduce the monitoring and intervention record: intervention labels restrict the allowed changes of coordinates, and passive operation excludes transformations that would produce negative intensities.

Only one positive recorded severity is needed to fix a labelled mechanism direction. The second measured level calibrates the relative severity response across recorded levels, as stated in Corollary 3, and is not required in the proof of the ambiguity set. Any equivalent explanation can be chosen as the reference. Changing the reference changes only the coordinates used to describe the ambiguity, not the set of equivalent signatures. Replacing \(A\) by \(AT_0^{-1}\) with \(T_0\in R_L^S\) replaces \(S\) by \(T_0S\). The equivalent signature set remains unchanged, and the new transformations are \(R_L^ST_0^{-1}\). Because \(z'_U=Nz_U\) holds for the invertible block \(N=T_{UU}\), LL2 also remains valid after changing the reference explanation.

\paragraph{Support monotonicity.}
Hold the reference coordinates and label constraints fixed. If \(S'\subseteq S\), then \(\Gamma_S\subseteq\Gamma_{S'}\) and \(R_L^S\subseteq R_L^{S'}\): a smaller passive support removes nonnegativity constraints and enlarges the ambiguity set. Two consequences must be kept apart. First, replacing the population support by a sampled subset \(S'\) is conservative in the following sense: any decision that is constant over \(R_L^{S'}\) is also constant over \(R_L^S\). Second, this conservatism concerns which explanations are allowed, not the states at which the decision is evaluated. Two explanations that agree at all sampled states may still disagree at an unobserved state. A valid check on sampled data therefore evaluates the decision over the enlarged set \(R_L^{S'}\), and it must do so at every state where the rule will be applied, not only at the sampled states. The decision may concern whether an intensity is increasing or whether it exceeds a maintenance threshold.

\section{Recovery of mechanism signatures and intensities}

Theorem 1 gives every explanation that the record cannot distinguish. This section identifies the quantities that remain unchanged across these explanations. Appendix A proves Corollaries 2 to 4 and Proposition 1.

\paragraph{Corollary 1 (block form and invariants).}
Under LL0, LL1 and LL2, ordering \(L\) before \(U\) gives

\begin{equation}
T=
\begin{bmatrix}
D&H\\
0&N
\end{bmatrix},
\qquad
D\in D_+^{|L|},
\quad
H\in\mathbb{R}^{|L|\times q},
\quad
N\in GL_q
\label{eq:block}
\end{equation}

with \(Dz_L+Hz_U\geq0\) and \(Nz_U\geq0\) on \(S\), and

\begin{equation}
A'_L=A_LD^{-1},
\qquad
A'_U=(A_U-A_LD^{-1}H)N^{-1},
\qquad
z'_L=Dz_L+Hz_U,
\qquad
z'_U=Nz_U.
\label{eq:blockinv}
\end{equation}

Thus, three quantities remain unchanged across all equivalent explanations. The full signature column space \(\mathrm{col}\,A\) remains fixed. Each labelled mechanism also keeps its own signature direction, up to a positive scale. After the labelled signature directions \(\mathrm{span}\,A_L\) are removed, the remaining subspace spanned by the unlabelled mechanisms is also fixed (see the complementary identity \(\mathrm{col}\,Q_LY^{\top}=\mathrm{col}\,Z_U^{\top}\) in Appendix A). A signature ray is a sensor response direction without fixed magnitude. Here, absorption means that part of an unlabelled mechanism contribution is reassigned to the intensity of a labelled mechanism without changing the total sensor signal. Labelled intensities may absorb such contributions, subject to Eq.~\eqref{eq:block}, but they cannot absorb another labelled intensity.

\paragraph{Proof.}
Preserving each named intervention direction forces zeros outside the corresponding diagonal entry, which gives Eq.~\eqref{eq:block}. Inverting this block matrix gives Eq.~\eqref{eq:blockinv}. Positive diagonal scaling changes only the magnitude of a labelled signature and does not change its direction. The invertible unlabelled block \(N\) may mix the unlabelled signatures, but it preserves their combined span after the labelled directions are removed. Therefore, the stated quantities remain the same across all equivalent explanations.

The block form separates three effects. These are rescaling of labelled mechanisms through \(D\), absorption of unlabelled contributions into labelled mechanisms through \(H\) and mixing among unlabelled mechanisms through \(N\). After the labelled signature directions are removed, let \(Q_L\) remove \(\mathrm{col}\,A_L\). For spanning noiseless signals \(Y=AZ\), the complementary identity is \(\mathrm{col}\,Q_LY^{\top}=\mathrm{col}\,Z_U^{\top}\). This shows that the remaining projected signal retains the combined information associated with the unlabelled mechanisms, even when the individual unlabelled signatures cannot be separated. Each labelled intensity row can still vary within an affine space parallel to \(\mathrm{row}\,Z_U\), which is determined by the unlabelled intensities. An affine space is a shifted linear space. Related results on complementarity and coupling were established by Sawall et al.~\cite{ref27} and Neymeyr and Sawall \cite{ref28}. Appendix A gives the corresponding identities without restrictions on signature signs. Related rotational restrictions were studied using binary shear \cite{ref26,ref27}, feasible solutions and dual Borgen plots \cite{ref29,ref30,ref31,ref32}, and partial identifiability \cite{ref33}.

\paragraph{Corollary 2 (dimension).}
Under LL0, LL1 and LL2,

\begin{equation}
\mathrm{dim}\,R_L^S
=
K^2-|L|(K-1)
=
K+(K-1)q.
\label{eq:dimension}
\end{equation}

Equation~\eqref{eq:dimension} gives the exact number of continuously adjustable parameters remaining after intervention labels are imposed. This complements earlier results on equality constraints \cite{ref25} and known zero restrictions in receptor modelling \cite{ref45}.

\paragraph{Corollary 3 (recorded severity response).}
Under LL0, LL1 and LL2, for \(k\in L\) and recorded severities \(l\geq0,l_*>0\), the ratio \(h_k(l)/h_k(l_*)\) is identifiable. Recorded responses are determined up to one positive scale factor. Between two recorded severities, the response is only constrained to lie between the corresponding neighbouring scaled values while remaining monotone. Beyond the recorded range, only monotonicity is imposed. Calibrating the shape of the response requires two or more distinct positive severities; the level \(l=0\) duplicates the healthy anchor and carries no response information.

\paragraph{Corollary 4 (unlabelled intensity).}
Under LL0, LL1 and LL2, if \(U=\{j\}\), then \(z'_j=\gamma z_j\), \(\gamma>0\). The preorder \(s\preccurlyeq_jt\Leftrightarrow s_j\leq t_j\), \(s,t\in S\), is preserved. Here, the preorder represents the ordering of passive states according to mechanism intensity. If \(q\geq2\), the intensity of a specific unlabelled mechanism cannot in general be recovered up to a positive scale on \(S\). However, this does not mean that every allowed transformation changes the ordering of passive states for that mechanism.

\begin{table}[!htbp]
\centering
\caption{Examples with three mechanisms under Theorem 1. Each additional intervention label removes part of the ambiguity. Stars denote free entries subject to invertibility and to the row constraints \(\Gamma_S\) imposed by joint passive occurrence. Matrix rows are separated by semicolons; \(d_1,d_2,d_3,\gamma>0\).}
\label{tab:labelsets}

{\footnotesize
\renewcommand{\arraystretch}{1.30}
\setlength{\extrarowheight}{2pt}

\begin{tabularx}{\textwidth}{@{}p{0.13\textwidth}p{0.23\textwidth}p{0.10\textwidth}X@{}}
\toprule
\textbf{Intervention labels \(L\)} &
\textbf{Transformation \(T\)} &
\textbf{Dimension} &
\textbf{Guaranteed recovery} \\
\midrule

\(\varnothing\) &
\(\begin{bmatrix} *&*&*\\ *&*&*\\ *&*&* \end{bmatrix}\) &
9 &
Signature span \(\mathrm{col}\,A\) \\[3.0em]

\(\{1\}\) &
\(\begin{bmatrix} d_1&*&*\\ 0&*&*\\ 0&*&* \end{bmatrix}\) &
7 &
Signature ray of mechanism 1 and the unlabelled signature span after removal of that ray \\[3.0em]

\(\{1,2\}\) &
\(\begin{bmatrix} d_1&0&*\\ 0&d_2&*\\ 0&0&\gamma \end{bmatrix}\) &
5 &
Labelled signature rays and mechanism 3 intensity up to positive scale \\[3.0em]

\(\{1,2,3\}\) &
\(\mathrm{diag}(d_1,d_2,d_3)\) &
3 &
Every signature and intensity up to positive scale \\[0.5em]

\bottomrule
\end{tabularx}
}
\end{table}

The dimensions 9, 7, 5 and 3 show that intervention labels progressively remove ambiguity. When all three mechanisms are labelled, only positive scaling of signatures and intensities remains.

To examine how unlabelled mechanisms may be absorbed into a labelled mechanism, fix a labelled mechanism \(k\) and a value \(\alpha>0\). For \(b\in\mathbb{R}^{q}\), define

\begin{equation}
\Phi_k(\alpha)
=
\{b:\alpha z_k+b^{\top}z_U\geq0\ \text{on }S\},
\qquad
\kappa_{kj}
=
\underset{z_j>0}{\mathrm{ess\,inf}}\,
\frac{z_k}{z_j},
\quad
j\in U.
\label{eq:phi}
\end{equation}

The essential infimum is the lowest ratio that matters under the pooled passive distribution, after states with zero probability are ignored. When every passive cell has positive weight, it is simply the infimum over \(S\cap\{z_j>0\}\). LL2 ensures that each ratio is defined for at least one relevant state. The ratios use all states in which the selected unlabelled mechanism is active. The other mechanisms do not need to be absent.

\paragraph{Proposition 1 (pairwise bounds).}
Under LL0, LL1 and LL2, with \(K=3\) and \(L=\{1\}\), separate pairwise bounds are not generally exact. There are occurrence patterns for which the bounds \(b_j\geq-\alpha\kappa_{1j}\) are not necessary for \(b\in\Phi_1(\alpha)\). There are also occurrence patterns for which the same pairwise bounds are not sufficient. A necessary condition must hold for every allowed decomposition. A sufficient condition guarantees that a decomposition is allowed. At \(\alpha=1\), examples are

\begin{equation}
S_N=
\{(1,1,0),(1,0,1),(0,1,1),(1,0,0)\},
\qquad
S_S=
\{(1,1,1),(1,1,0),(1,0,1),(1,0,0)\}.
\label{eq:occurrence}
\end{equation}

\begin{figure}[!htbp]
\centering
\includegraphics[width=0.86\linewidth]{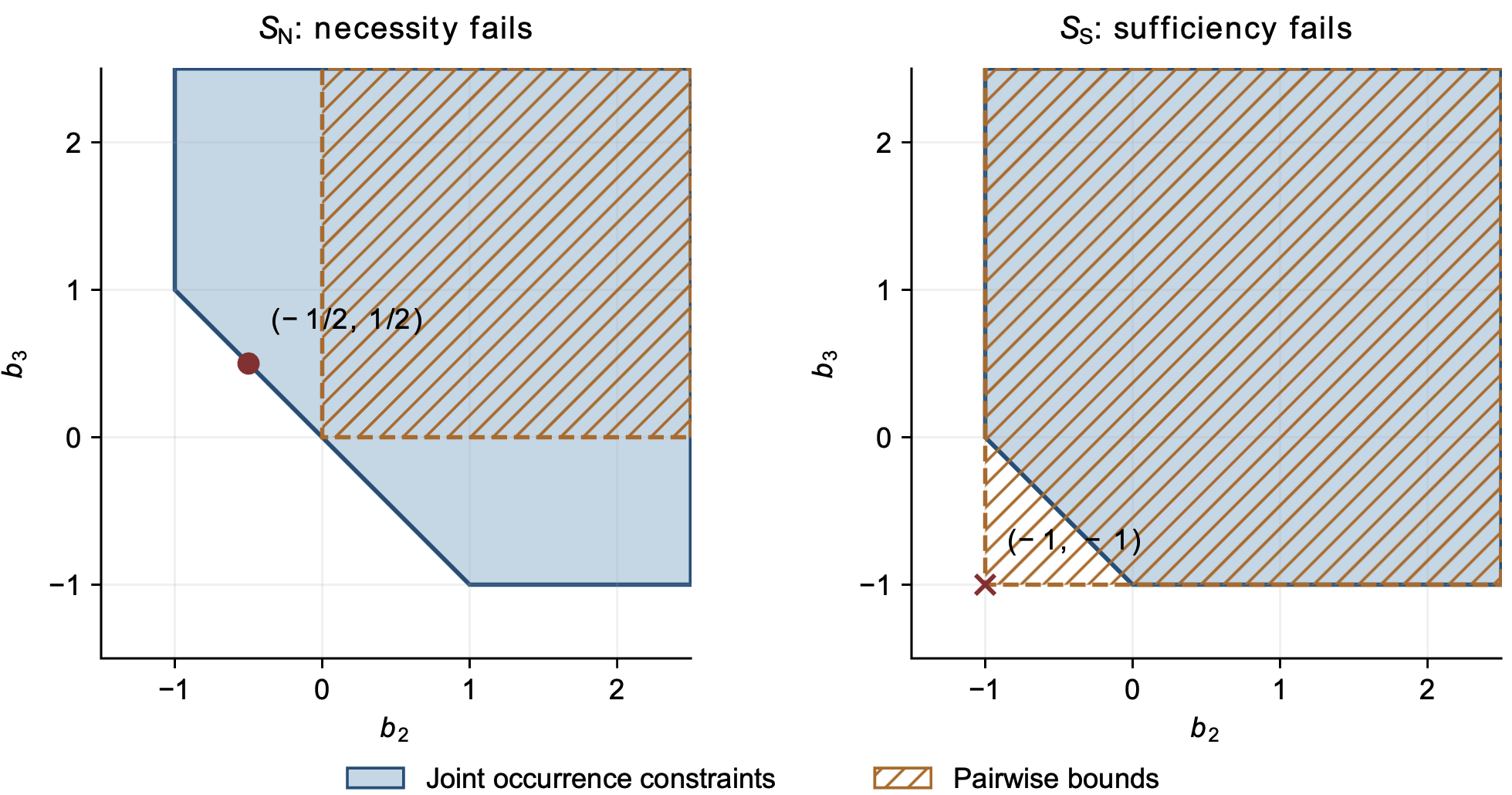}
\caption{Exact absorption regions \(\Phi_1\) at \(\alpha=1\) for the occurrence patterns \(S_N\) (left) and \(S_S\) (right), clipped to the displayed window. Dashed lines show the separate pairwise bounds. Under \(S_N\) the pairwise bounds exclude the feasible point \((-1/2,1/2)\). Under \(S_S\) they admit the infeasible point \((-1,-1)\).}
\label{fig:pairwise}
\end{figure}

Figure~\ref{fig:pairwise} gives the exact feasible regions. When several mechanisms occur together, the allowed value of one absorption coefficient can depend on the others. Therefore, the feasible region generally cannot be obtained from separate pairwise intervals \cite{ref23,ref24}.

\paragraph{Remark 1 (group envelope).}
The ambiguity set is not necessarily closed under matrix multiplication or inversion. The smallest transformation group that contains it is called the group envelope and is given by

\begin{equation}
\left\langle R_L^S\right\rangle
=
\begin{cases}
\mathcal{T}_L, & q\geq2,\\
\mathcal{T}_L\cap\{T_{jj}>0\}, & U=\{j\},\\
D_+^K, & q=0.
\end{cases}
\label{eq:group-envelope}
\end{equation}

A transvection is a simple shear transformation that adds a multiple of one coordinate to another. Positive diagonal matrices, permutations among the unlabelled mechanisms \(U\) and the stated nonnegative transvections belong to the ambiguity set. Their inverses generate signed transvections and lead to Eq.~\eqref{eq:group-envelope}. The group envelope can contain transformations that are not allowed by the ambiguity set because they may violate the passive occurrence constraints.

If the reference additionally satisfies \(a_k\notin\mathrm{col}\,B\) for every \(k\), the restricted equivalence set removes members whose unlabelled column of \(AT^{-1}\) lies in \(\mathrm{col}\,B\). This excludes an unlabelled mechanism signature that could be represented entirely by the declared operating condition basis. This restriction is a physical-plausibility screen on reference parameterizations, not an identifiability constraint, because \(B\) is already fixed by Lemma A.1. Theorem 1 does not impose it.

\section{Example with two mechanisms and sensor calibration}

Independent information outside the mixed monitoring signal can further restrict the remaining ambiguity. The following example shows this for two mechanisms.

\paragraph{Example 1 (independent sensor calibration).}
Under LL0, LL1 and LL2, let \(K=2\) and \(L=\{1\}\). Hats denote one fixed reference explanation. Write \(a_2(t)=\widehat{a}_2-t\widehat{a}_1\) for the mechanism-2 signature obtained from Eq.~\eqref{eq:blockinv}, after the positive scale of the unlabelled mechanism is fixed. Here \(t=T_{12}/T_{11}\) is the single remaining shear coefficient, and Corollary 1 gives \(t\geq-\kappa_{12}\). Changing \(t\) moves part of the unlabelled contribution into the labelled mechanism while leaving the total sensor signal unchanged.

A sensor is indexed by \(c\in\{1,\ldots,p\}\); for a vector \(a\), \(a_c\) denotes its entry at sensor \(c\). The reference sensor \(c_0\) is independently verified as unaffected by mechanism 1, so \(\widehat{a}_{1,c_0}=0\). Suppose also \(\widehat{a}_{2,c_0}\neq0\) and \(\widehat{a}_{1,c}\neq0\). Independent calibration must come from information outside the mixed decomposition. It supplies a known bound \(\rho_c\geq0\) on the mechanism-2 signature at sensor \(c\) relative to sensor \(c_0\): \(|a_2(t)_c|\leq\rho_c|a_2(t)_{c_0}|\). Define \(\mathcal{M}_{\rho}\) as the members of \(\mathcal{M}\) that satisfy this bound. Physically, the bound limits the allowed value of the mechanism-2 signature at sensor \(c\) relative to its value at the reference sensor \(c_0\). The bound is deterministic and defines which explanations are allowed. It is not a statistical confidence interval.

Because \(\widehat{a}_{1,c_0}=0\), the entry at the reference sensor does not change with \(t\): \(a_2(t)_{c_0}=\widehat{a}_{2,c_0}\). The calibration bound is therefore the condition \(|\widehat{a}_{2,c}-t\widehat{a}_{1,c}|\leq\rho_c|\widehat{a}_{2,c_0}|\). This condition holds exactly when \(t\in[\overline{t}-\Delta,\overline{t}+\Delta]\), with \(\overline{t}=\widehat{a}_{2,c}/\widehat{a}_{1,c}\) and \(\Delta=\rho_c|\widehat{a}_{2,c_0}/\widehat{a}_{1,c}|\). The passive-occurrence constraint adds \(t\geq-\kappa_{12}\). The shear coefficients compatible with both the monitoring record and the independent sensor calibration are therefore exactly

\begin{equation}
F
=
[-\kappa_{12},\infty)
\cap
[\overline{t}-\Delta,\overline{t}+\Delta],
\qquad
\overline{t}
=
\frac{\widehat{a}_{2,c}}{\widehat{a}_{1,c}},
\qquad
\Delta
=
\rho_c
\left|
\frac{\widehat{a}_{2,c_0}}{\widehat{a}_{1,c}}
\right|.
\label{eq:F}
\end{equation}

\begin{figure}[!htbp]
\centering
\includegraphics[width=0.86\linewidth]{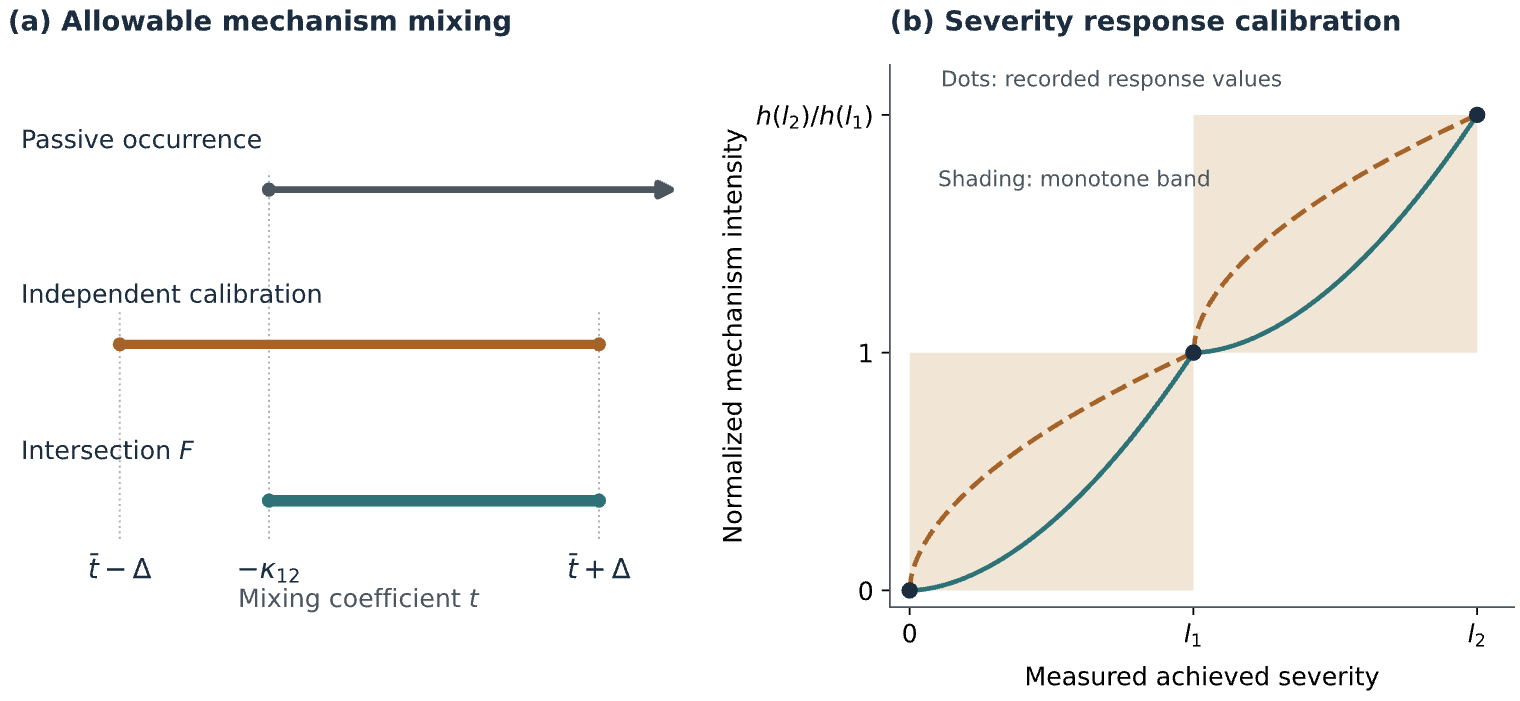}
\caption{Two distinct roles of calibration. Panel (a) shows how independent sensor calibration restricts the mechanism mixing coefficient to an interval while passive occurrence imposes a lower bound. Their intersection gives the allowed set in Example 1. The displayed case has \(\overline{t}-\Delta<-\kappa_{12}<\overline{t}+\Delta\). Panel (b) shows how recorded intervention severities fix the normalized severity response at the measured levels. Any strictly increasing response between consecutive levels remains allowed. The shaded regions show deterministic ranges allowed by the model. They are not statistical confidence intervals.}
\label{fig:calibration}
\end{figure}

The numerical values of the cone \(\Gamma_S\), the ratios \(\kappa_{kj}\) and the shear set \(F\) depend on the chosen reference explanation. The complete set of explanations that reproduce the observations does not depend on this choice. Sensor calibration alone permits the calibration interval in Eq.~\eqref{eq:F}, while passive occurrence imposes the lower bound in the same equation. Equation~\eqref{eq:F} is the intersection of these two restrictions, as shown in Figure~\ref{fig:calibration}(a). If \(F\) is nonempty, the maximum of \(|a_2(t)_c|/|a_2(t)_{c_0}|\) over \(t\in F\) equals \(\rho_c\) and is attained at the upper endpoint \(t=\overline{t}+\Delta\), which gives Eq.~\eqref{eq:calmax}.

\begin{equation}
\max_{t\in F}
\frac{|a_2(t)_c|}
{|a_2(t)_{c_0}|}
=
\rho_c.
\label{eq:calmax}
\end{equation}

The passive boundary \(-\kappa_{12}\) reaches the calibration bound only when it coincides with an endpoint of the calibration interval. If \(F\) has positive length, this occurs when \(-\kappa_{12}=\overline{t}-\Delta\). If \(-\kappa_{12}=\overline{t}+\Delta\), the feasible set contains only one point. An empty intersection means that no explanation in \(\mathcal{M}_{\rho}\) satisfies both the monitoring record and the independent calibration.

Equation~\eqref{eq:F} is related to the feasible boundary analysis of Lawton and Sylvestre \cite{ref19}. Their coefficient restriction is similar to \(\kappa_{12}\), because both require the reconstructed coefficients to remain nonnegative. Lawton and Sylvestre use an unknown resolving transformation, so their coordinates differ from those used here \cite{ref19}. In the present formulation, mechanism signatures \(A\) may contain positive or negative values. Therefore, the spectral nonnegativity restriction in Lawton and Sylvestre has no direct equivalent here. After the labelled direction is fixed, the remaining occurrence restriction is the half line in Eq.~\eqref{eq:F}. The uniqueness condition in Lawton and Sylvestre instead depends on complementary zeros in two nonnegative factors.

\section{Implications for diagnosis and test design}

Structural fault isolability \cite{ref12,ref13} asks whether a given model structure and sensor set allow residuals to distinguish faults; it presumes an identified model and is usually stated for one fault at a time. Decision sufficiency, defined below, is a complementary requirement on the record itself. A mechanism intensity may be unrecoverable while a specific threshold decision is still the same for every explanation in the ambiguity set. Conversely, a design that isolates single faults can fail decision sufficiency when labels are partial and mechanisms occur together. In the terminology of partial identification, the record identifies a set of decompositions rather than a single one; a decision is justified exactly when it is constant over this identified set \cite{ref46,ref47}.

The practical value of the ambiguity analysis is that a maintenance decision does not require every mechanism parameter to be identified uniquely.

\paragraph{Definition 1 (decision sufficiency).}
Fix a maintenance rule that returns a decision, for example inspect or do not inspect, for any explanation \((A',z',h')\) allowed by the record. The record is decision sufficient for this rule if every explanation in the ambiguity set of Theorem 1, under every monotone extension of the severity responses permitted by LL1, returns the same decision.

In other words, the underlying mechanism decomposition need not be unique if all explanations consistent with the record lead to the same decision. The decision may concern only what is needed in practice, such as whether an intensity is increasing or whether it exceeds a threshold linked to a recorded severity. Corollary 3 links the intensity of a labelled mechanism to physical severity at the recorded intervention levels. Calibrating the severity response at the recorded levels does not, however, uniquely determine the passive intensity. The record is decision sufficient for a threshold rule only when the same outcome is obtained under all mixing allowed by Corollary 1.

Consider the case with two mechanisms from Example 1. Cooler fouling is labelled as mechanism 1 and filter clogging remains unlabelled. Set \(\widehat{a}_1=(8,0)^{\top}\), \(\widehat{a}_2=(1,1)^{\top}\), and \(S=\{z_a=(2,1),z_b=(1,3)\}\). Two distinct positive intervention severities \(l_a<l_*\) have responses \(h_1(l_a)=1\) and \(h_1(l_*)=2\) in reference intensity units. The ratio \(\kappa_{12}=\min\{2/1,1/3\}=1/3\) is the smallest ratio \(z_1/z_2\) over the passive states with \(z_2>0\), so every \(t\geq-1/3\) is allowed. With \(T_{11}=\alpha_1\) and \(T_{12}=\alpha_1t\), Eq.~\eqref{eq:blockinv} gives \(z'_1(z_b)=\alpha_1(1+3t)\) and \(h'_1(l_*)=2\alpha_1\). The normalized value \(z'_1(z_b)/h'_1(l_*)=(1+3t)/2\) therefore does not depend on \(\alpha_1\). The cooler inspection threshold \(z'_1(z_b)>h'_1(l_*)\) holds exactly when \(t>1/3\). The allowed explanation \(t=0\) gives no inspection, whereas the allowed explanation \(t=1\) gives inspection. Both explanations reproduce the same monitoring record. Therefore, the record is not decision sufficient for this threshold.

Corollary 4 gives a different result for filter clogging. The increase in clogging intensity from \(z_a\) to \(z_b\) is preserved under every allowed transformation. Thus, the clogging trend is recoverable even though the cooler inspection decision is not.

When independent calibration additionally supplies \(\rho_c=1\), with \(c\) the first sensor and \(c_0\) the second, Eq.~\eqref{eq:F} gives \(F=[0,1/4]\). Over this interval, the maximum of \((1+3t)/2\) is \(7/8<1\), so the threshold is not exceeded for any allowed shear coefficient. The mechanism decomposition therefore remains nonunique, but the maintenance decision becomes unique.

Corollary 2 and Table~\ref{tab:labelsets} show how each intervention label removes \(K-1\) free parameters, giving \(9\rightarrow7\rightarrow5\rightarrow3\) for three mechanisms. When \(q\geq2\), Corollary 1 still identifies the labelled signature directions and, after these directions are removed, the combined subspace spanned by the unlabelled signatures. Thus, the record identifies the part of the degradation signal associated with the unlabelled mechanisms as a group, even when that contribution cannot be assigned to individual unlabelled mechanisms.

If the diagnosis requires an intensity trend for a specific mechanism, Corollary 4 gives a stricter requirement. At most one mechanism can remain unlabelled for that trend to be guaranteed under the stated assumptions. Proposition 1 also shows that the amount of unlabelled contribution that can be absorbed into a labelled mechanism depends on which mechanisms occur together during passive operation.

In the case with two mechanisms, a finite set of passive states with \(z_2>0\) can give a larger minimum ratio than the population value \(\kappa_{12}\). This produces a larger ambiguity set. If the same decision is obtained for every explanation in this larger set, the decision must also be the same throughout the population ambiguity set. In this sense, the check is conservative. As stated with support monotonicity in Section 3, the conservatism concerns which explanations are allowed; the decision must still be evaluated at every state where the rule will be applied.

These results give a direct principle for acquisition design. First, specify the decision that the monitoring system must support for the mechanism of interest. Then determine whether the same decision is obtained for every explanation allowed by the planned interventions and passive operating conditions. If the decision changes across these explanations, more information is required. This information may come from another labelled intervention, broader passive occurrence or independent calibration.

\subsection{A linear-programming check for decision sufficiency}

Let \(S'=\{z^{(1)},\ldots,z^{(n)}\}\) be the sampled passive support in reference coordinates and let \(Z^*\) be the states at which the rule will be applied. Consider a threshold rule on a labelled mechanism \(k\). The decision compares \(z'_k\) with \(h'_k(l_*)\), and both scale with the same factor \(\alpha_k\), so the decision depends only on the normalized absorption vector \(b=(T_{kj}/\alpha_k)\), \(j\in U\).

\paragraph{Step 1.}
Form the polyhedron
\[
P=
\left\{
b:
z_k^{(i)}+b^{\top}z_U^{(i)}\geq0
\ \text{for } i=1,\ldots,n
\right\}
\cap C,
\]
where \(C\) is the calibration interval of Eq.~\eqref{eq:F} when available and \(C=\mathbb{R}^{q}\) otherwise.

\paragraph{Step 2.}
For each \(z^*\) in \(Z^*\), minimize and maximize the linear objective
\[
\frac{z_k^*+b^{\top}z_U^*}{\widehat{h}_k(l_*)}
\]
over \(b\in P\). These are two linear programs.

\paragraph{Step 3.}
The record is decision sufficient at \(z^*\) exactly when both optima lie on the same side of the threshold. For a trend decision between states \(z_a\) and \(z_b\), use the objective
\[
z_{b,k}-z_{a,k}
+
b^{\top}(z_{b,U}-z_{a,U})
\]
and check that its sign is the same at both optima.

Only row \(k\) of \(T\) enters \(z'_k\). Any admissible row can be completed to a full member of the ambiguity set, for example with identity blocks elsewhere, which leave all other intensities unchanged and therefore nonnegative. The polyhedron \(P\) therefore describes exactly the attainable values of \(b\). By support monotonicity the check is conservative with respect to the allowed explanations, but it must be run at every intended application state \(z^*\), not only at the sampled states.

\section{Limitations and conclusions}

Theorem 1 is an identifiability result at the population level. LL0, LL1 and LL2 and the corresponding population supports are assumed to be known exactly. Therefore, the theory does not provide uncertainty bounds for finite samples, guarantees of numerical stability near rank deficiency or statistical tests for the acquisition assumptions.

Constant signatures and common noise can be restrictive when system responses change with operating condition. Applying the theory separately to different operating regimes requires LL0, LL1 and LL2 to be checked again in each regime. This approach does not by itself solve cases in which the signatures depend on operating condition or the noise variance changes with operating condition. Regime-wise application also multiplies the acquisition effort: every regime needs its own healthy anchors in each analysed session, its own labelled interventions with verified absences, and its own spanning passive variation, so with \(R\) regimes the intervention burden grows roughly by the factor \(R\). A possible extension is to let each signature vary linearly with the declared condition basis; this enlarges the parameter set, strengthens the required spanning condition, and is left to future work.

The number of mechanisms \(K\) is treated as known. Under LL0, LL1 and LL2 it equals the dimension of the space spanned by the deconvolved degradation signals \(Az\) over all cells, provided \(p\geq K\). Underestimating \(K\) merges mechanisms and invalidates the verified-absence constraints. Overestimating \(K\) introduces directions that LL2 cannot anchor. Rank selection under noise is outside the population scope of this paper.

The fixed-direction constraint \(Te_k=\alpha_ke_k\) rests on exact absence verification. If a nontarget mechanism were active at small intensity during an intervention, this constraint would relax to a narrow cone, and the exact equalities of Theorem 1 would become approximate: the true ambiguity set would then lie between a slightly smaller and a slightly larger set. A perturbation analysis is left to future work, and Table~\ref{tab:acquisition} lists the corresponding verification duty.

Without sufficient LL2 spanning, alternative signatures can leave the reference column space when \(p>K\). In that case, the transformations in Theorem 1 may not include all possible explanations.

Corollary 3 constrains the severity response only at and between the recorded levels. Interventions involving several mechanisms at the same time and signatures that depend on operating condition require further theory. The analytical examples do not estimate failure probability, optimize maintenance, establish safe operation or provide empirical validation.

Within these assumptions, Theorem 1 gives the complete set of transformations that preserve the monitoring record. The corollaries identify the mechanism quantities that remain unchanged and show how intervention labels reduce ambiguity. Proposition 1 shows why mechanisms that occur at the same time must be considered jointly.

The main perspective of this paper is that condition monitoring should not rely only on one fitted decomposition. Instead, the analysis should determine which conclusions remain the same across all explanations consistent with the monitoring and intervention record. A decision concerning a specific mechanism is justified when the same decision is obtained across the ambiguity set, even when the underlying decomposition is not unique. Maritime machinery is used as the running example, but the framework applies to engineering condition monitoring problems that satisfy the stated assumptions. The main design question is therefore whether the acquisition itself provides enough information for the intended decision.


\section*{Appendix A. Nuisance identification and proofs of the consequences}

\paragraph{Lemma A.1 (nuisance identification).}
Under LL0, any two models in \(\mathcal{M}\) that generate the same cellwise sensor distributions share \(B\), the session offsets \(c_u\) and the law of \(\varepsilon\), and hence the cellwise distributions of \(Az\).

\paragraph{Proof.}
In a healthy anchor, \(x=B\,g(w)+c_u+\varepsilon\). With the lower median of each noise coordinate set to zero, the coordinatewise lower median of \(x\) equals \(B\,g(w)+c_u\). Differences between anchors of the same session cancel \(c_u\). Because these differences span \(\mathbb{R}^{d}\) by LL0(b), they determine \(B\). One healthy anchor per session then determines \(c_u\). After \(B\,g(w)+c_u\) is removed, the anchor residuals give the law of \(\varepsilon\). Since the characteristic function of \(\varepsilon\) is nonzero everywhere, the deconvolution of every cell is unique and gives the law of \(Az\).

\paragraph{Complementary space and coupling identities without sign restrictions.}
This proof shows that removing the labelled signature directions does not remove any independent combination of the unlabelled signatures. It also describes the remaining freedom in the labelled intensities. After removing the labelled sensor directions, the projected relation is \(Q_LY=CZ_U\), where \(C=Q_LA_U\).

Suppose that a nonzero combination of the unlabelled signatures disappeared after this projection. If \(Cv=0\), then \(A_Uv\) would lie in \(\mathrm{col}\,A_L\). This would contradict the full column rank of \(A\) unless \(v=0\). Therefore, \(C\) has full column rank and \(Z_U=C^{+}Q_LY\), which gives \(\mathrm{row}\,Q_LY=\mathrm{row}\,Z_U\).

For normalized \(A_L\), let \(V=A_L^{+}Y\). Since \(V=Z_L+A_L^{+}A_UZ_U\), each labelled intensity row satisfies \(Z_{L,i}\in V_i+\mathrm{row}\,Z_U\). An affine space is a shifted linear space. Conversely, every simultaneous choice within these affine spaces can be obtained by replacing \(A_U\) with \(A_U-A_LH\) and \(Z_L\) with \(Z_L+HZ_U\). These transformations change how the intensity is divided among the mechanisms while preserving full rank and the sensor signal product. Nonnegativity then restricts the allowed choices through Eq.~\eqref{eq:block}. If a block is empty, the corresponding condition does not apply. No restriction on the signs of the mechanism signatures is required.

\paragraph{Corollary 2.}
Each labelled mechanism prevents mixing into the other \(K-1\) entries of its transformation column. This leaves \(K^2-|L|(K-1)\) free entries. It remains to show that the nonnegativity condition during passive operation does not reduce this dimension. Let \(Q\) contain ones in every allowed position outside the diagonal and zeros elsewhere. For a sufficiently small positive value of \(\epsilon>0\), \(I+\epsilon Q\) is invertible and positive in every allowed position. The same properties continue to hold for matrices sufficiently close to this constructed matrix within the constrained transformation space. These matrices map \(S\geq0\) to nonnegative intensities and therefore belong to \(R_L^S\). Hence, the ambiguity set has the full dimension of the constrained transformation space. Substitution of \(|L|=K-q\) gives Eq.~\eqref{eq:dimension}.

\paragraph{Corollary 3.}
Every recorded response of a labelled mechanism receives the same positive multiplier \(\alpha_k\) under Theorem 1. Dividing by the response at \(l_*\) removes the unknown scale. Between recorded severities, strict monotonicity limits the response to values between the neighbouring recorded responses. Any strictly intermediate value can be obtained with an increasing interpolation. An increasing extension can also be defined beyond the largest recorded severity. Unobserved severities can therefore be assigned consistently with the transformed intensities. They do not further restrict the severity response beyond the recorded levels and the monotonicity requirement.

\paragraph{Corollary 4.}
With one unlabelled mechanism, Eq.~\eqref{eq:block} gives the alternative intensity as \(z'_j=\gamma z_j\). Invertibility excludes \(\gamma=0\). LL2 provides at least one state with \(z_j>0\). Nonnegativity therefore requires \(\gamma>0\). All intensity comparisons are consequently preserved. When \(q\geq2\), choose two distinct indices \(i,j\in U\). The permutation that exchanges these two unlabelled mechanisms preserves the monitoring record and nonnegativity throughout the passive support \(S\). Suppose that \(z'_i=z_j\) were a positive multiple of \(z_i\) for every passive state in \(S\). This proportionality would force the projected unlabelled intensities to satisfy a nonzero linear relation, contradicting LL2. Therefore, when more than one mechanism remains unlabelled, the intensity of a particular unlabelled mechanism cannot generally be recovered up to a positive scale. However, this does not mean that every allowed exchange changes the ordering of passive states for that mechanism.

\paragraph{Proposition 1.}
The two occurrence patterns provide separate counterexamples to necessity and sufficiency. Both contain the projected states \((1,0)\) and \((0,1)\) and therefore satisfy LL2. For the first occurrence pattern \(S_N\), \(\kappa_{12}=\kappa_{13}=0\). Requiring the labelled intensity to remain nonnegative at every state gives \(b_2\geq-1\), \(b_3\geq-1\), and \(b_2+b_3\geq0\). The displayed coefficient pair \((-1/2,1/2)\) violates one of the separate pairwise bounds. However, it still gives the nonnegative intensities \(1/2,3/2,0,1\). Therefore, the pairwise bounds are not necessary. For the second occurrence pattern \(S_S\), both ratios equal one and the exact inequalities are \(b_2\geq-1\), \(b_3\geq-1\), and \(b_2+b_3\geq-1\). The displayed coefficient pair \((-1,-1)\) satisfies the separate pairwise bounds. However, it gives a negative intensity \(-1\) at \((1,1,1)\). Therefore, the pairwise bounds are not sufficient. Each example defines a complete invertible transformation using \(D=1\), \(N=I\), and \(H=(b_2,b_3)\). Full rank signatures, fixed nuisance terms and the prescribed intervention cells then define models satisfying LL0 and LL1. The stated occurrence patterns satisfy LL2.


\section*{Acknowledgements}

This work was supported by the Research Council of Finland under the Academy Research Fellowship project TRANSITION-MODEL [grant no.~362947].

\section*{CRediT authorship contribution statement}

\textbf{Ahmad BahooToroody:}
Conceptualization, Methodology, Formal analysis, Writing -- original draft, Writing -- review \& editing.

\textbf{Mohammad Mahdi Abaei:}
Conceptualization, Methodology, Formal analysis, Writing -- original draft, Writing -- review \& editing.



\begin{thebibliography}{99}

\bibitem{ref1}
Rinehart AW, Simon DL.
An integrated architecture for aircraft engine performance monitoring and fault diagnostics: engine test results.
In: Proceedings of the 50th AIAA/ASME/SAE/ASEE Joint Propulsion Conference, Propulsion and Energy Forum; 2014.
doi:10.2514/6.2014-3924.

\bibitem{ref2}
Agarwal V, Lybeck N, Bickford R, Rusaw R.
Development of asset fault signatures for prognostic and health management in the nuclear industry.
In: 2014 IEEE Conference on Prognostics and Health Management (PHM); 2014. p.~1--7.
doi:10.1109/ICPHM.2014.7036366.

\bibitem{ref3}
Coraddu A, Oneto L, Ghio A, Savio S, Anguita D, Figari M.
Machine learning approaches for improving condition-based maintenance of naval propulsion plants.
Proc Inst Mech Eng Part M 2016;230(1):136--153.
doi:10.1177/1475090214540874.

\bibitem{ref4}
Zio E.
Prognostics and Health Management (PHM): Where are we and where do we (need to) go in theory and practice.
Reliab Eng Syst Saf 2022;218:108119.
doi:10.1016/j.ress.2021.108119.

\bibitem{ref5}
Song C, Liu K.
Statistical degradation modeling and prognostics of multiple sensor signals via data fusion: A composite health index approach.
IISE Trans 2018;50(10):853--867.
doi:10.1080/24725854.2018.1440673.

\bibitem{ref6}
Fang X, Yan H, Gebraeel N, Paynabar K.
Multi-sensor prognostics modeling for applications with highly incomplete signals.
IISE Trans 2021;53(5):597--613.
doi:10.1080/24725854.2020.1789779.

\bibitem{ref7}
Coraddu A, Oneto L, Ghio A, Savio S, Anguita D, Figari M.
Condition based maintenance of naval propulsion plants [data set].
UCI Machine Learning Repository 2014.
doi:10.24432/C5K31K.

\bibitem{ref8}
Chehade A, Bonk S, Liu K.
Sensory-based failure threshold estimation for remaining useful life prediction.
IEEE Trans Reliab 2017;66(3):939--949.
doi:10.1109/TR.2017.2695119.

\bibitem{ref9}
Wang H, Li R, Tang G, Yuan H, Zhao Q, Cao X.
A compound fault diagnosis for rolling bearings method based on blind source separation and ensemble empirical mode decomposition.
PLoS One 2014;9(10):e109166.
doi:10.1371/journal.pone.0109166.

\bibitem{ref10}
Wang M, Zhang W, Shao M, Wang G.
Separation and extraction of compound-fault signal based on multi-constraint non-negative matrix factorization.
Entropy 2024;26:583.
doi:10.3390/e26070583.

\bibitem{ref11}
Hao L, Gebraeel N, Shi J.
Simultaneous signal separation and prognostics of multi-component systems: the case of identical components.
IIE Trans 2015;47(5):487--504.
doi:10.1080/0740817X.2014.955357.

\bibitem{ref12}
Murata R, Marzat J, Piet-Lahanier H, Boujnah S, Belleoud P.
Fault detection and isolation based on structural analysis: Application to a multi-engine propulsion cluster.
Sensors 2025;25(4):1054.
doi:10.3390/s25041054.

\bibitem{ref13}
Düştegör D, Frisk E, Cocquempot V, Krysander M, Staroswiecki M.
Structural analysis of fault isolability in the DAMADICS benchmark.
Control Eng Pract 2006;14(6):597--608.
doi:10.1016/j.conengprac.2005.04.008.

\bibitem{ref14}
Fu K, Disanayaka Mudiyanselage SS, Dai C, Kim M.
Prognostics of multisensor systems with unknown and unlabeled failure modes via Bayesian nonparametric process mixtures.
arXiv:2602.19263v1; 2026. Preprint.

\bibitem{ref15}
Tsiatis A.
A nonidentifiability aspect of the problem of competing risks.
Proc Natl Acad Sci USA 1975;72:20--22.
doi:10.1073/pnas.72.1.20.

\bibitem{ref16}
Bedford T, Lindqvist BH.
The identifiability problem for repairable systems subject to competing risks.
Adv Appl Probab 2004;36(3):774--790.
doi:10.1239/aap/1093962233.

\bibitem{ref17}
Massonis G, Villaverde AF.
Finding and breaking Lie symmetries: implications for structural identifiability and observability in biological modelling.
Symmetry 2020;12(3):469.
doi:10.3390/sym12030469.

\bibitem{ref18}
Massonis G, Banga JR, Villaverde AF.
AutoRepar: a method to obtain identifiable and observable reparameterizations of dynamic models with mechanistic insights.
Int J Robust Nonlinear Control 2023;33(9):5039--5057.
doi:10.1002/rnc.5887.

\bibitem{ref19}
Lawton WH, Sylvestre EA.
Self modeling curve resolution.
Technometrics 1971;13:617--633.
doi:10.1080/00401706.1971.10488823.

\bibitem{ref20}
Tauler R, Smilde A, Kowalski B.
Selectivity, local rank, three-way data analysis and ambiguity in multivariate curve resolution.
J Chemom 1995;9:31--58.
doi:10.1002/cem.1180090105.

\bibitem{ref21}
Borgen OS, Kowalski BR.
An extension of the multivariate component-resolution method to three components.
Anal Chim Acta 1985;174:1--26.
doi:10.1016/S0003-2670(00)84361-5.

\bibitem{ref22}
Rajkó R.
Natural duality in minimal constrained self modeling curve resolution.
J Chemom 2006;20(3--4):164--169.
doi:10.1002/cem.999.

\bibitem{ref23}
Golshan A, Abdollahi H, Maeder M.
Resolution of rotational ambiguity for three-component systems.
Anal Chem 2011;83(3):836--841.
doi:10.1021/ac102429q.

\bibitem{ref24}
Golshan A, Abdollahi H, Beyramysoltan S, Maeder M, Neymeyr K, Rajkó R, Sawall M, Tauler R.
A review of recent methods for the determination of ranges of feasible solutions resulting from soft modelling analyses of multivariate data.
Anal Chim Acta 2016;911:1--13.
doi:10.1016/j.aca.2016.01.011.

\bibitem{ref25}
Beyramysoltan S, Rajkó R, Abdollahi H.
Investigation of the equality constraint effect on the reduction of the rotational ambiguity in three-component system using a novel grid search method.
Anal Chim Acta 2013;791:25--35.
doi:10.1016/j.aca.2013.06.043.

\bibitem{ref26}
Manne R.
On the resolution problem in hyphenated chromatography.
Chemom Intell Lab Syst 1995;27:89--94.
doi:10.1016/0169-7439(95)80009-X.

\bibitem{ref27}
Sawall M, Fischer C, Heller D, Neymeyr K.
Reduction of the rotational ambiguity of curve resolution techniques under partial knowledge of the factors. Complementarity and coupling theorems.
J Chemom 2012;26:526--537.
doi:10.1002/cem.2458.

\bibitem{ref28}
Neymeyr K, Sawall M.
On an SVD-free approach to the complementarity and coupling theory: A note on the elimination of unknowns in sums of dyadic products.
J Chemom 2016;30:30--36.
doi:10.1002/cem.2765.

\bibitem{ref29}
Sawall M, Neymeyr K.
On the area of feasible solutions and its reduction by the complementarity theorem.
Anal Chim Acta 2014;828:17--26.
doi:10.1016/j.aca.2014.04.026.

\bibitem{ref30}
Sawall M, Jürß A, Schröder H, Neymeyr K.
Simultaneous construction of dual Borgen plots. I: The case of noise-free data.
J Chemom 2017;31:e2954.
doi:10.1002/cem.2954.

\bibitem{ref31}
Sawall M, Moog A, Kubis C, Schröder H, Selent D, Franke R, Brächer A, Börner A, Neymeyr K.
Simultaneous construction of dual Borgen plots. II: Algorithmic enhancement for applications to noisy spectral data.
J Chemom 2018;32:e3012.
doi:10.1002/cem.3012.

\bibitem{ref32}
Neymeyr K, Sawall M.
On the set of solutions of the nonnegative matrix factorization problem.
SIAM J Matrix Anal Appl 2018;39(2):1049--1069.
doi:10.1137/17M1118439.

\bibitem{ref33}
Gillis N, Rajkó R.
Partial identifiability for nonnegative matrix factorization.
SIAM J Matrix Anal Appl 2023;44:27--52.
doi:10.1137/22M1507553.

\bibitem{ref34}
Donoho D, Stodden V.
When does non-negative matrix factorization give a correct decomposition into parts?
In: Advances in Neural Information Processing Systems 16. 2003. NeurIPS proceedings.

\bibitem{ref35}
Huang K, Sidiropoulos ND, Swami A.
Non-negative matrix factorization revisited: uniqueness and algorithm for symmetric decomposition.
IEEE Trans Signal Process 2014;62(1):211--224.
doi:10.1109/TSP.2013.2285514.

\bibitem{ref36}
Ding C, Li T, Jordan MI.
Convex and semi-nonnegative matrix factorizations.
IEEE Trans Pattern Anal Mach Intell 2010;32:45--55.
doi:10.1109/TPAMI.2008.277.

\bibitem{ref37}
Walter E, Pronzato L.
Qualitative and quantitative experiment design for phenomenological models---A survey.
Automatica 1990;26(2):195--213.
doi:10.1016/0005-1098(90)90116-Y.

\bibitem{ref38}
Sawall M, Kubis C, Schröder H, Meinhardt D, Selent D, Franke R, Brächer A, Börner A, Neymeyr K.
Multivariate curve resolution methods and the design of experiments.
J Chemom 2020;34(2):e3159.
doi:10.1002/cem.3159.

\bibitem{ref39}
Escobar LA, Meeker WQ.
Planning accelerated life tests with two or more experimental factors.
Technometrics 1995;37(4):411--427.
doi:10.1080/00401706.1995.10484374.

\bibitem{ref40}
Meeker WQ, Escobar LA.
Pitfalls of accelerated testing.
IEEE Trans Reliab 1998;47(2):114--118.
doi:10.1109/24.722271.

\bibitem{ref41}
Buchholz S, Rajendran G, Rosenfeld E, Aragam B, Schölkopf B, Ravikumar P.
Learning linear causal representations from interventions under general nonlinear mixing.
Advances in Neural Information Processing Systems 2023;36.
Author preprint, arXiv:2306.02235v2.

\bibitem{ref42}
Squires C, Seigal A, Bhate SS, Uhler C.
Linear causal disentanglement via interventions.
In: Proceedings of the 40th International Conference on Machine Learning.
PMLR 2023;202:32540--32560. Proceedings.

\bibitem{ref43}
Ahuja K, Mahajan D, Wang Y, Bengio Y.
Interventional causal representation learning.
In: Proceedings of the 40th International Conference on Machine Learning.
PMLR 2023;202:372--407. Proceedings.

\bibitem{ref44}
Lorber A, Faber K, Kowalski BR.
Net analyte signal calculation in multivariate calibration.
Anal Chem 1997;69(8):1620--1626.
doi:10.1021/ac960862b.

\bibitem{ref45}
Park ES, Spiegelman CH, Henry RC.
Bilinear estimation of pollution source profiles and amounts by using multivariate receptor models.
Environmetrics 2002;13(7):775--798.
doi:10.1002/env.557.

\bibitem{ref46}
Manski CF.
Partial Identification of Probability Distributions.
New York: Springer; 2003.

\bibitem{ref47}
Tamer E.
Partial identification in econometrics.
Annu Rev Econ 2010;2:167--195.
doi:10.1146/annurev.economics.050708.143401.

\end{thebibliography}
\end{document}